\documentclass[twocolumn,aps,amsmath,showpacs]{revtex4}
\usepackage{graphicx}
\newcommand{\be}{\begin{equation}}
\newcommand{\ee}{\end{equation}}

\newcommand{\bea}{\begin{eqnarray}}
\newcommand{\eea}{\end{eqnarray}}
\newcommand{\bd}{\begin{displaymath}}
\newcommand{\ed}{\end{displaymath}}
\newcommand{\bi}{\begin{itemize}}
\newcommand{\ei}{\end{itemize}}
\newcommand{\bc}{\begin{center}}
\newcommand{\ec}{\end{center}}
\newcommand{\bfl}{\begin{flushleft}}
\newcommand{\efl}{\end{flushleft}}
\newcommand{\bfr}{\begin{flushright}}
\newcommand{\efr}{\end{flushright}}
\newcommand{\f}{\frac}

\def\bk{{\bf k}}

\def\6{\partial}  
  \def\ve{\varepsilon}

\def\={\!\!\!&=&\!\!\!}
\def\+{\!\!\!&&\!\!\!+~}
\def\-{\!\!\!&&\!\!\!-~}

\begin{document}
\author{A. L. Monteros$^1$, G. S. Uppal$^1$, S. R. McMillan$^1$,  M. Crisan$^{2}$, and I. \c{T}ifrea$^{1}$}

\affiliation{$^1$Department of Physics, California State University, Fullerton, CA 92834, USA}
\affiliation{$^2$Department of Physics, ``Babe\c{s}-Bolyai" University, 40084 Cluj-Napoca, Romania}

\date{\today}
\title{Thermoelectric transport properties of a T-shaped double quantum dot system in the Coulomb blockade regime}


\begin{abstract}
We investigate the thermoelectric properties of a T-shaped double quantum dot system described by a generalized Anderson Hamiltonian. The system's electrical conduction (G) and the fundamental thermoelectric parameters such as the Seebeck coefficient ($S$) and the thermal conductivity ($\kappa$), along with the system's thermoelectric figure of merit (ZT) are numerically estimated based on a Green's function formalism that includes contributions up to the Hartree-Fock level. Our results account for finite onsite Coulomb interaction terms in both component quantum dots and discuss various ways leading to an enhanced thermoelectric figure of merit for the system. We demonstrate that the presence of Fano resonances in the Coulomb blockade regime is responsible for a strong violation of the Wiedemann-Franz law and a considerable enhancement of the system's figure of merit ($ZT$).
\end{abstract}
\pacs{72.60.+g,73.63.Kv,73.50.Lw}
\maketitle

\section{Introduction}

Efficient thermoelectric materials are the key ingredient for feasible energy conversion technology. The standard measure for thermal efficient materials is their so called figure of merit ($ZT$), a parameter that combines information about the material charge and thermal conduction, $ZT=G S^2 T/\kappa$.  Here, $T$ is the temperature, $G$ is the electrical conductivity, $S$ is the Seebeck coefficient, and $\kappa$ is the total thermal conductivity (contributions to the thermal conductivity are due to both the electronic and phonon systems, $\kappa=\kappa_e+\kappa_p$). For practical applications, $ZT$ has to be as large as possible, meaning that we are looking for materials characterized by a large Seebeck coefficient and electrical conductivity at the same time with a small thermal conductivity. In conventional solids the ratio $\kappa_e/G T$ is constant as they obey the well known Wiedemann-Franz law \cite{ashcroft}, and usually  an increase in the electric conductivity will reduce the Seebeck coefficient \cite{mott}. Accordingly, for conventional solids the figure of merit is relatively low, $ZT<1$. 

About 20 years ago Hicks and Dresselhaus \cite{hicks1,hicks2} suggested that systems with low dimensionality such as multi-layer quantum well superlattices or one dimensional quantum wires can be the ideal candidates for efficient thermoelectric devices. According to their calculations, when the system's dimensionality is reduced, $ZT$ can increase by more than an order of magnitude \cite{hicks1,hicks2}. In general, the lower the dimensionality of the system, the higher the figure of merit. The calculations are not relevant only for semiconductors, the material can be a metal, semiconductor, or semimetal as long as the system's dimensionality is reduced to two or one dimensions. Additionally, when the system's dimensionality is reduced, there will be an increase in the phonon scattering from the material's surfaces, leading to a decrease in the phonon thermal conductivity, $\kappa_p$. A smaller thermal conductivity will further enhance the system's figure of merit \cite{venka,harman,hochbaum,boukai,yoshida}.

Single quantum dot or multiple quantum dot systems are other low dimensional systems available for practical applications based on their electric and thermal properties \cite{davies}.  The physics associated with quantum dot systems is a result of strong quantum confinement and strong onsite Coulomb interaction.  For example, single quantum dot system's are subject to the Kondo effect when the spin of the electron on the localized levels of the quantum dot gets screened below a characteristic temperature (the Kondo temperature) \cite{glazman}. Also, electron transport in single quantum dot systems is characterized by Fano-like resonances attributed to the interference between the quantum amplitudes for the two possible electron channels in the system: a direct conduction pathway through the leads and an indirect conduction pathway that involves the localized levels in the quantum dot \cite{sasaki}.

The situation is somehow different in multiple quantum dot systems where due to additional conduction channels one can observe more complex signatures of the  Kondo and Fano effects \cite{bulka,aharony,sasaki,kobayashi,orellana}. In particular, in a two quantum dot system with one dot connected to the external leads and the other to the first dot but not to the external leads (T-shaped double-quantum-dot system), one has to account for a two state Kondo effect, as the localized spins in the system's component quantum dots are screened at different temperatures  \cite{chung,zitko,crisan}. In T-shaped double-quantum-dot systems the Fano effect is related to the presence of two electron transport channels, one involving only the dot coupled directly to the external leads and the other involving both the central and side dots \cite{guevara,trocha0}. The Fano effect in the double quantum dot system is attributed to different strength couplings of these two transport channels to the external leads \cite{trocha0}. The Kondo and Fano effects in quantum dot systems are directly observed in electron transport experimental data. On the other hand, we expect that the variety of possible physics phenomena in quantum dot systems will strongly influence the system's thermal properties, leading to the possible violation of the Wiedemann-Franz law or of the Mott relation. Indeed, signatures of the Kondo and Fano effects are also identified in experimental and theoretical studies of thermoelectric properties of quantum dot systems and molecules (See the review articles \cite{wang,dubi}).

Here, we consider the thermoelectric properties of a T-shaped double quantum dot  system. As the name suggests, the system contains two quantum dots arranged in a T-shaped configuration, with the main dot connected to external leads and the side dot connected to the main dot but not to the external leads. The system is characterized by two possible conduction channels (one involving only the main dot, and the other involving both the main and the side dots) and as a consequence of multiple scattering and interference of transmitted waves through both channels  a Fano resonance associated with a Kondo resonance can be observed \cite{kobayashi,sasaki}. Several authors considered double quantum dot systems in connection with thermoelectric properties. Liu and Fang \cite{liu} analyzed a double-coupled-quantum-dot molecule sandwiched between two metallic electrodes and found an enhancement of the figure of merit when the system is in the Fano resonance regime.  Liu's and Fang's analysis includes effects related to a magnetic flux, but they completely ignore the onsite Coulomb interaction effects \cite{liu}. Karlstrom {\em et al.} arrive at the same conclusion when they considered a two-level system in connection with possible electron wave interference \cite{karlstrom}. Trocha and Barnas considered a double quantum dot system with both system's quantum dots connected to external magnetic/nonmagnetic leads \cite{trocha}. Their investigation of the system's thermoelectric properties is based on the Hartree-Fock approximation and included effects due to the onsite Coulomb interaction terms (for simplicity the system's two component dots were considered to be identical) \cite{trocha}. According to their results, the thermoelectric efficiency of the system can be increased  due to interference effects and Coulomb interactions \cite{trocha}. As we already mentioned, our system's configuration is slightly different than the one considered by Trocha and Barnas, the system's component dots are different and only the main dot is connected to external leads.  Our results are relevant for temperatures higher than the specific temperature associated to the first stage Kondo effect \cite{chung,zitko,crisan}, a regime that it is well described by the Hartree-Fock approximation.

The paper is organized as follows. Section II presents our model and the general formalism used to investigate thermoelectric properties in quantum dot systems. In Section III we present our numerical results. Finally, Section IV includes a summary of our results and conclusions.

\section{Thermoelectric Properties}

The T-shaped double quantum dot system consists of two quantum dots, a main dot connected to external leads and a side dot connected to the main dot but not to the external leads. Each dot is characterized by a localized electron level, $\ve_i$ ($i=1$ for the main dot and $i=2$ for the side dot). To describe the system's properties we use a generalized Anderson Hamiltonian \cite{anderson} with free electrons provided by the external leads and the localized impurity levels identified by the localized electron levels in each component quantum dot:

\begin{eqnarray}\label{hamiltonian}
H&=&\sum_{\bk,\sigma;\alpha}\ve_\bk c^\dagger_{\bk\sigma;\alpha}c_{\bk\sigma;\alpha} +\sum_{i=1}^2\ve_i
\sum_\sigma a^\dagger_{i\sigma} a_{i\sigma}\nonumber\\
&+&\sum_{i=1}^2 U_i n_{i\sigma}n_{i-\sigma}+\sum_{\bk\sigma;\alpha}V_{\bk 1;\alpha}\left(c^\dagger_{\bk\sigma;\alpha}a_{1\sigma}+
a^\dagger_{1\sigma}c_{\bk\sigma;\alpha}\right)\nonumber\\
&+&t\sum_\sigma\left(a^\dagger_{1\sigma}a_{2\sigma}+
a^\dagger_{2\sigma}a_{1\sigma}\right)\;.
\end{eqnarray}
Above, the free electrons in the external leads are described by the Hamiltonian's first term; $c^\dagger_{\bk\sigma;\alpha}$ and $c_{\bk\sigma;\alpha}$ are the fermionic creation and annihilation operators for electrons with momentum $\bk$ and spin $\sigma$ in the lead $\alpha$ ($\alpha\equiv$ left (L), right (R)). 
Localized  electrons in the main dot ($i=1$) and the side dot ($i=2$) have characteristic energies $\ve_i$ and they are subject to local Coulomb interactions of local strength $U_i$ ($a^\dagger_{i\sigma}$ and $a_{i\sigma}$ are fermionic creation and annihilation operators for localized electrons with spin $\sigma$ in the $i$'th quantum dot, and $n_{i\sigma}=a^\dagger_{i\sigma}a_{i\sigma}$ is the number of particle operator corresponding to the electronic level $\ve_i$ and electron spin $\sigma$). Electrons provided by the external leads and electrons localized in the main dot interact via an interaction term of strength  $V_{\bk 1;\alpha}$. For simplicity we will consider the case $V_{\bk 1;L}=V_{\bk 1;R}$ in which the detector couples to the leads only in the symmetric combination $c_{\bk\sigma}=(c_{\bk\sigma;L}+c_{\bk\sigma;R})/\sqrt{2}$ and the dot connects effectively to a single lead, with $V_{\bk 1}=\sqrt{2}V_{\bk 1;L}$. Note that there is no interaction between electrons in the side dot and the free electrons in the leads. Finally, localized electrons in the two component dots interact via a hopping-like term of strength $t$.

The thermoelectric response of a system can be understood relatively simply. The electrical conductivity $G$ is a measure of how easily the electrical charges flow through the system to generate a voltage drop.  On the other hand, the Seebeck coefficient $S$ measures the voltage induced in a sample subject to a temperature gradient. Finally, the thermal conductivity $\kappa$ measures how heat can be transferred through the system to maintain a temperature gradient. These coefficients combine into the system's figure of merit $ZT=G S^2 T/\kappa$ to measure the total thermoelectric efficiency of a material. The transport coefficients can be obtained within the linear response theory. We can introduce a general function \cite{kim}
\begin{equation}
L_{n\sigma}=-\f{1}{h}\int d\ve (\ve-\mu)^n\f{\partial f}{\partial \ve} T_\sigma(\ve)\;,
\end{equation}
where $\mu$ is the chemical potential, $f(\ve)$ is the Fermi-Dirac distribution function, $T_\sigma(\ve)$ is the system's transmission coefficient, and $h$ is the Planck constant. We can express all thermoelectric coefficients as functions of the general function $L_{n\sigma}$ \cite{trocha,kim}
\begin{equation}\label{conductivity}
G=e^2\sum_\sigma L_{0\sigma}\;,
\end {equation}
\begin{equation}\label{seebeck}
S=-\f{1}{eT}\f{\sum_\sigma L_{1\sigma}}{\sum_\sigma L_{0\sigma}}\;,
\end{equation}
\begin{equation}\label{thermalconductance}
\kappa_e=\f{1}{T}\left[\sum_\sigma L_{2\sigma}-\f{\left(\sum_\sigma L_{1\sigma}\right)^2}{\sum_\sigma L_{0\sigma}}\right]\;.
\end{equation}
The system's transmission coefficient, $T_\sigma(\ve)$, can be calculated using the electronic Green's function that describes electron-electron correlations in the main dot, $G_{11}^\sigma(\omega)=\left<\left<a_{1\sigma};a^\dagger_{1\sigma}\right>\right>$:
\begin{equation}\label{transmission}
T_\sigma(\ve)=-\Delta {\textrm Im} G^\sigma_{11}(\ve)\;,
\end{equation}
where $\Delta=\pi \sum_\bk |V_{1\bk}|^2\delta(\ve-\ve_\bk)$. The calculation of the electronic correlation functions can be performed using the equation of motion (EOM) method \cite{hewson}. The EOM method involves a relatively simple technique that allows the evaluation of the main dot electronic Green's function starting form the system's Hamiltonian (see Eq. (\ref{hamiltonian})). The technique generates at each step higher order correlation functions, so, when a two electron correlation function is considered (as the one we need for the evaluation of the transmission coefficient) the EOM method will connect this function to a four electron correlation function that has to be evaluated using the same method. The procedure will lead to an infinite hierarchy of higher order correlation functions that eventually has to be truncated using an appropriate approximation. This complication is manly due to the onsite Coulomb interaction terms present in the system's Hamiltonian. Within the Hartree-Fock approximation, the main dot Green's function can be extracted exactly as \cite{hewson,tifrea1,tifrea2}
\begin{eqnarray}\label{G11}
&&G_{11}^\sigma(\omega)=\nonumber\\
&&\f{1}{\f{\left(\omega-\ve_1\right)\left(\omega-\ve_1-U_1\right)}{\omega-\ve_1-U_1\left(1-\left<n_{1-\sigma}\right>\right)}-t^2\f{\omega-\ve_1-U_2\left(1-\left<n_{2-\sigma}\right>\right)}{\left(\omega-\ve_2\right)\left(\omega-\ve_2-U_2\right)}-\Sigma_\bk^0(\omega)}\;\;\;,\nonumber\\
\end{eqnarray}
where 
\begin{equation}
\Sigma_\bk^0(\omega)=\sum_\bk\f{|V_{\bk 1}|^2}{\omega-\ve_\bk}\;.
\end{equation}
Although the equation for the main dot Green's function is exact within the Hartree-Fock approximation, $G_{11}^\sigma(\omega)$ depends on the average occupation of both the main and the side dots, $\left<n_{1-\sigma}\right>$ and $\left<n_{2-\sigma}\right>$. This dependence makes the problem self-consistent as the average occupation number in a quantum dot is related to the electron-electron correlation function
\begin{equation}\label{avgnumber}
\left<n_{i-\sigma}\right>=-\f{1}{\pi}\int f(\omega) {\textrm Im} G^{-\sigma}_{ii} (\omega) d\omega\;.
\end{equation}
Additionally, to compute the electron-electron correlation function in the main dot we need to know the average occupation number in the side dot and implicitly the electron-electron correlation function in this dot. 
Using the same approximation one finds:
\begin{eqnarray}\label{G22}
&&G_{22}^\sigma(\omega)=\nonumber\\
&&\f{1}{\f{\left(\omega-\ve_2\right)\left(\omega-\ve_2-U_2\right)}{\omega-\ve_2-U_2\left(1-\left<n_{2-\sigma}\right>\right)}-\f{t^2}{\f{\left(\omega-\ve_1\right)\left(\omega-\ve_1-U_1\right)}{\omega-\ve_1-U_1\left(1-\left<n_{1-\sigma}\right>\right)}-\Sigma_\bk^0(\omega)}}\;\;.\nonumber\\
\end{eqnarray}
Eqs. (\ref{G11}), (\ref{avgnumber}), and (\ref{G22}) allow the evaluation of the system's transmission function and thereafter an estimation of the system's thermoelectric properties. Despite the exact analytical formulas for the electronic Green's functions in both component quantum dots, the equations are self-consistent and their final solutions require numerical calculations.

\section{Numerical results}

\begin{figure}[t]
\centering \scalebox{1.5}[1.5]{\includegraphics*{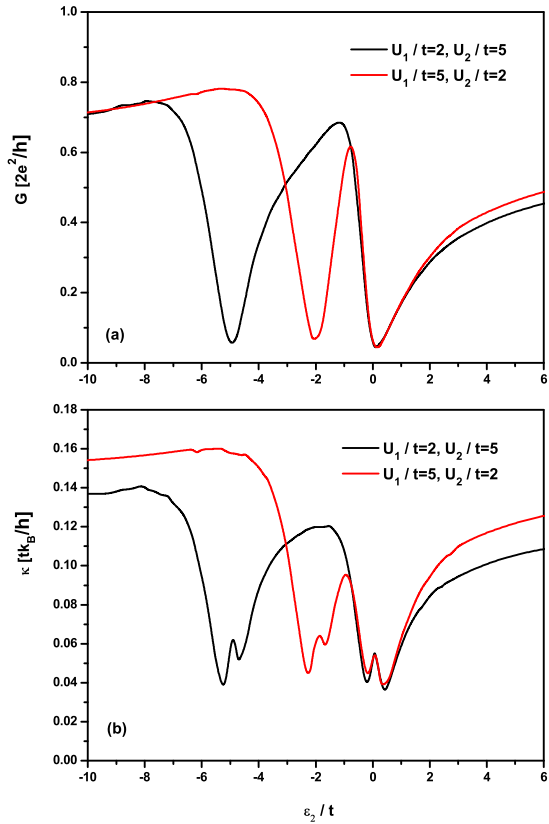}}
\caption{(Color online) The system's electron conductivity (a)  and thermal conductivity (b) as function of the energy level in the side dot for different values of the onsite Coulomb interaction ($U_1/t=2$, $U_2/t=5$ - black line; $U_1/t=5$, $U_2/t=2$ - red line). Other parameters in the system are $\ve_1 /t =-0.2$, $\Delta/t=0.5$, $k_B T/t=0.1$, and $\ve_F /t =0$.}
\label{fig1}
\end{figure}

For numerical calculations it is convenient to introduce dimensionless quantities and measure all energies in units of the inter-dot coupling constant $t$ relative to the Fermi level of the conduction electrons in the leads ($\ve_F=0$). We consider the most general case with finite onsite Coulomb interaction in both the main and side dots of the system ($U_1\neq U_2$). Our numerical analysis is performed as function of the energy level in the system's side dot, $\ve_2/t$. Unless otherwise mentioned, we consider the following values in our calculations $\ve_1/t=-0.2$, $\Delta/t=0.5$, and $k_B T/t=0.1$. In particular, we consider the slow detection limit ($\Delta/t<1$) as in this case the presence of a side dot in the system strongly influences its conduction properties \cite{tifrea1}. 

Figure \ref{fig1} presents results for the system's electrical and thermal conductivities as function of the side dot energy for various values of the onsite Coulomb interaction in the main and side dots ($U_1/t=2$, $U_2/t=5$ - black line; $U_1/t=5$, $U_2/t=2$ - red line). The system's electrical conductivity (Figure \ref{fig1}a) presents two dips associated with the Fano effect, one at $\ve_2=\ve_F=0$ and the other one at $\ve_2=-U_2$. As we discussed, this behavior is consistent with the existence of two possible conduction channels in the system. On the other hand, the system's thermal conductivity (Figure \ref{fig1}b) presents a richer structure in particular at energies close to the ones at which the two Fano dips occur in the electrical conductivity. Consequently, it is clear that in the slow detection limit, the Wiedemann-Franz law is  violated in the case of a T-shaped double quantum dot  system.

\begin{figure}[t]
\centering \scalebox{1.8}[1.8]{\includegraphics*{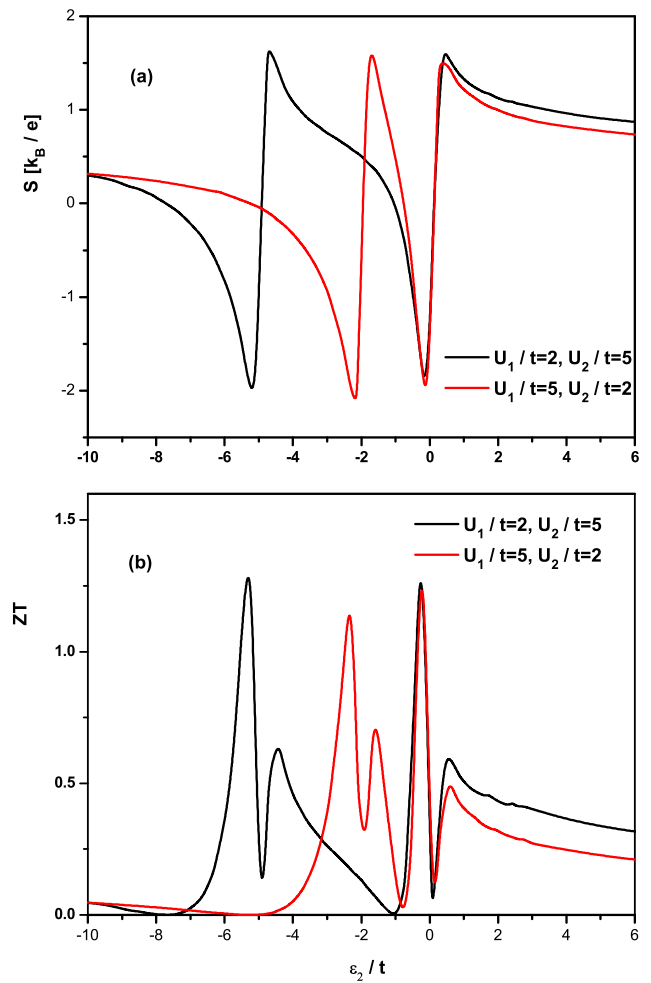}}
\caption{(Color online) The system's Seebeck coefficient (a)  and figure of merit (b) as function of the energy level in the side dot for different values of the onsite Coulomb interaction ($U_1/t=2$, $U_2/t=5$ - black line; $U_1/t=5$, $U_2/t=2$ - red line). Other parameters for the system are $\ve_1 /t =-0.2$, $\Delta/t=0.5$, $k_B T/t=0.1$, and $\ve_F /t =0$.}
\label{fig2}
\end{figure}

Figure \ref{fig2} presents results for the system's Seebeck coefficient and figure of merit for the same parameters as in Figure \ref{fig1}. The system's Seebeck coefficient (Figure \ref{fig2}a) presents a similar structure as the electrical and thermal conduction, with dips at energy values corresponding to the Fano resonances. In particular, the sign of the Seebeck coefficient changes, as the system's current switches from electrons (positive Seebeck coefficient) to holes (negative Seebeck coefficient). As mentioned in Ref. \cite{trocha}, when $\ve_2$ corresponds to one of the relevant resonances, a compensation between the charge current due to electrons by that due to holes will lead to no net current and no net voltage drop in the system and consequently to a vanishing Seebeck coefficient.

The rich structure of the electrical and thermal conductivities and of the Seebeck coefficient in the vicinity of the Fano resonances reflects also on the system's figure of merit $ZT$. Figure \ref{fig2}b presents the system's figure of merit as function of the energy level in the side dot (same parameters as before). Although the figure of merit is relatively small for most of the values of the energy in the side dot, near both resonances associated with the Fano effect, $ZT$ is considerably enhanced, with maximum values close to 2 for the particular values we considered.

In general, different values of the onsite Coulomb interaction in the main and side quantum dots of the system do not reflect in drastic changes of the system's thermoelectric properties. The structure of all considered parameters stays the same, the only difference being the position of the resonance related to the value of the onsite Coulomb interaction in the side dot. When $U_2$ increases, this resonance is pushed towards lower values of $\ve_2$ and it is not possible to be detected in experiments. Accordingly, all system's parameters will present only a single resonance structure. Another difference is the relative width and depth of the two resonances, however there is no major qualitative change when the values of the onsite Coulomb interaction terms vary.

\begin{figure}[t]
\centering \scalebox{0.95}[0.95]{\includegraphics*{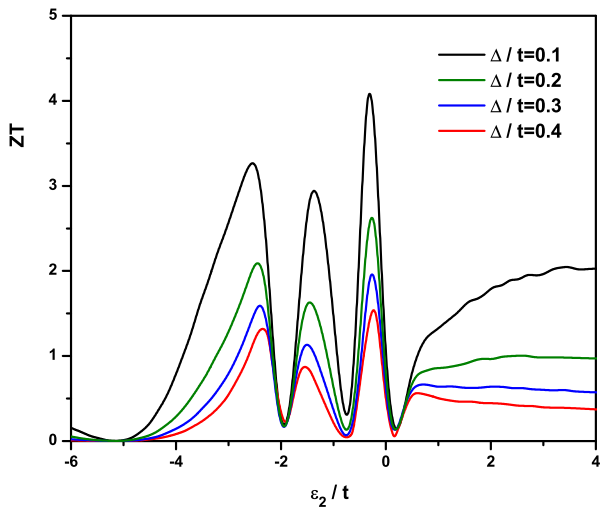}}
\caption{(Color online) The system's figure of merit  as function of the energy level in the side dot for different values of coupling between free electrons in the leads and localized electrons in the main dot ( $\Delta/t=0.1$ - black line; $\Delta/t=0.2$ - green line; $\Delta/t=0.3$ - blue line; $\Delta/t=0.4$ - redline). Other parameters for the system are $\ve_1 /t =-0.2$, $U_1/t=5$, $U_2/t=2$, $k_B T/t=0.1$, and $\ve_F /t =0$.}
\label{fig3}
\end{figure}

Consider now in more detail the slow detector limit of the system. One possibility to tune the system's thermoelectric properties is to vary the coupling between the conduction electrons in the leads and the localized electrons in the main dot ($\Delta/t$) \cite{zheng}. As one can see from Figure \ref{fig1}, in the case of the T-shaped double quantum dot  system, the Wiedemann-Franz law is violated, meaning that the electrical and thermal conductivity of the system behave differently. This situation also reflects on the value of $ZT$. We expect that with the reduction of $\Delta$, both the electric and thermal conductivities will decrease, however, this can result in a large enhancement of the system's figure of merit if the changes in the two parameters are not related to each other. Figure \ref{fig3} presents the system's figure of merit for various values of the ratio $\Delta/t$ for the case $U_1/t=5$ and $U_2/t=2$. As the coupling between the conduction electrons in the leads and the localized electrons in the main dot decreases, the value of the system's figure of merit increases near both resonances associated with the Fano effect. A value as high as $ZT\simeq 4$ can be predicted for $\Delta/t=0.1$. 

\begin{figure}[t]
\centering \scalebox{0.55}[0.55]{\includegraphics*{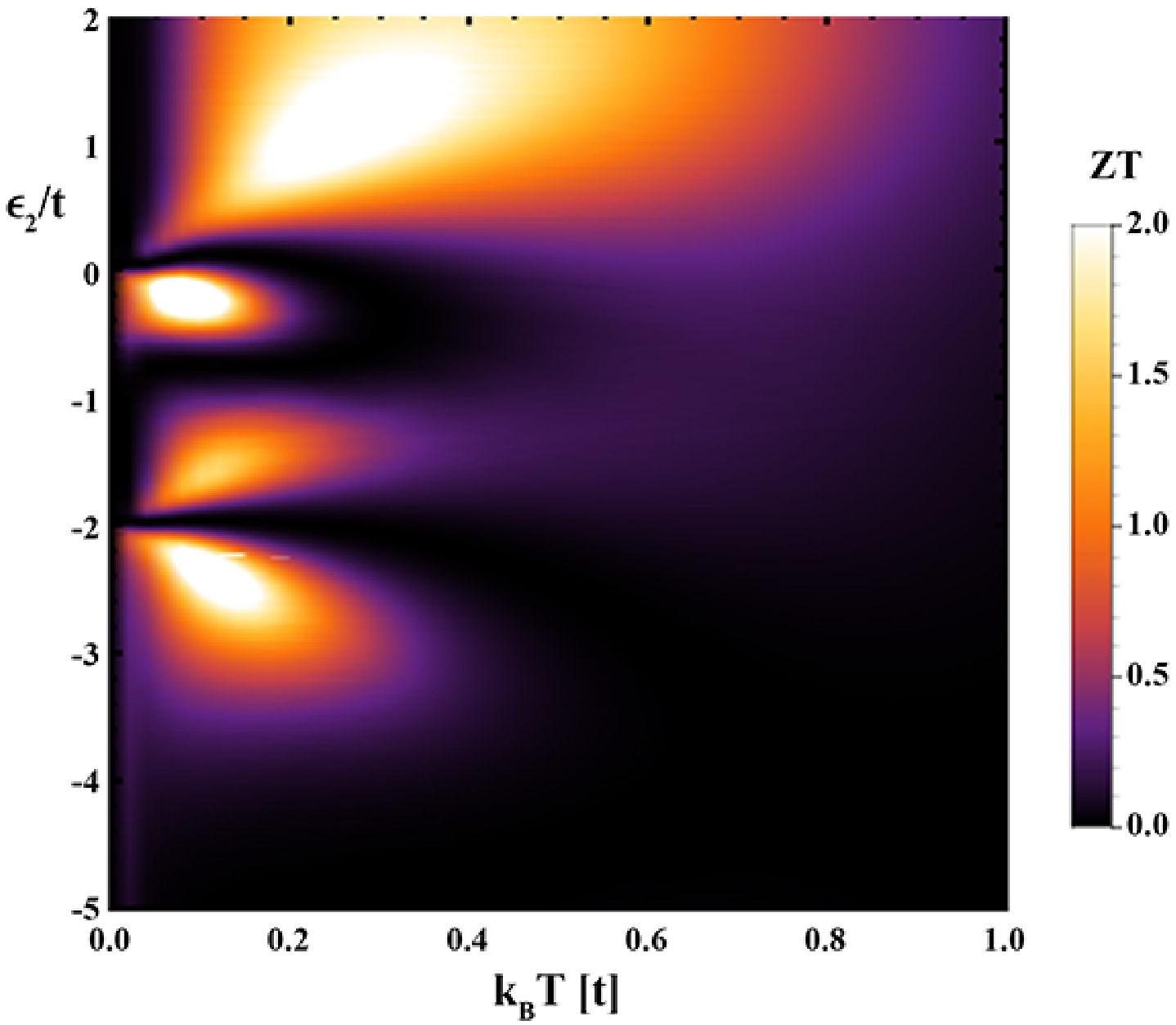}}
\centering \scalebox{0.95}[0.95]{\includegraphics*{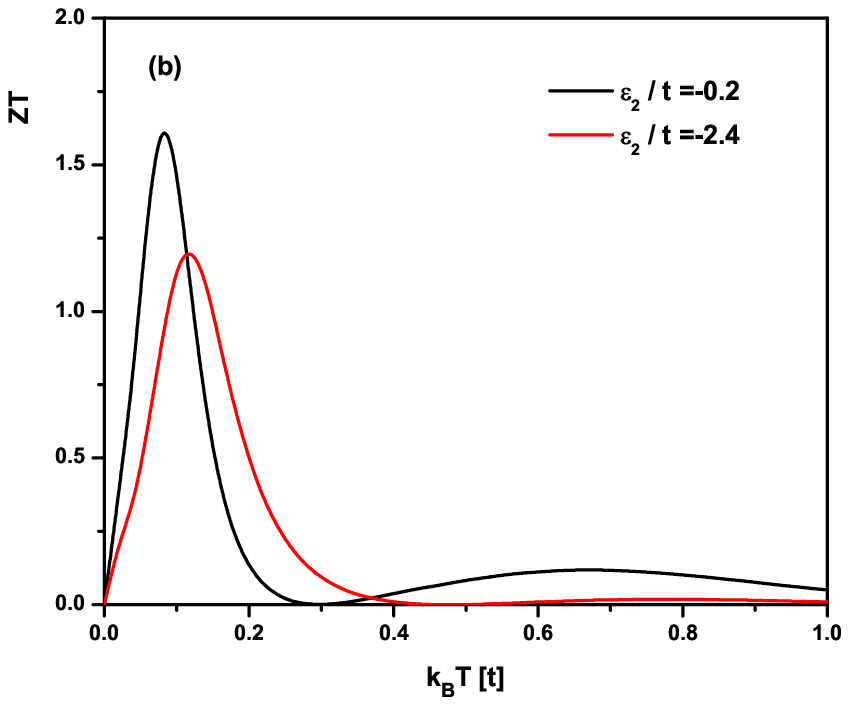}}
\caption{(Color online) (a) The system's figure of merit  as function of temperature and energy level in the side dot. (b) The system's figure of merit as function of temperature in the vicinity of the two Fano resonances ($\ve_2/t=-0.2$ - black line and $\ve_2/t=-2.4$ - red line). Other parameters for the system are $\ve_1 /t =-0.2$, $U_1/t=5$, $U_2/t=2$, $\Delta/t=0.5$ , and $\ve_F /t =0$.}
\label{fig4}
\end{figure}

Finally, consider the temperature dependence of the system's figure of merit, $ZT$. Figure \ref{fig4}a presents the figure of merit $ZT$ as function of temperature, $k_B T$ (units of the inter dot interaction $t$) and energy level in the side dot, $\ve_2/t$, for $U_1/t=5$, $U_2/t=2$, and $\Delta/t=0.5$. The system's figure of merit is enhanced in the vicinity of the two Fano resonances, and it presents a maximum at temperatures of the order of $k_BT/t\sim 0.1$. In Fig. \ref{fig4}b, we plot the temperature dependence of the figure of merit for  two selected  values of the side dot energy level around the two Fano resonances, $\ve_2/t=-0.2$ and $\ve_2/t=-2.4$. The temperature dependence of the system's figure of merit is similar for both cases, however, its maximum is situated at slightly different temperature values. Trocha and Barnas reported a high value for the figure of merit in the case of a double quantum dot system with both dots coupled to external leads ($ZT\sim 300$) for the simple situation without onsite Coulomb interaction  ($U_1=U_2=0$) \cite{trocha}. In our case, the finite onsite Coulomb interaction terms affect the structure of the figure of merit dependence on temperature and energy level in the side dot: several additional islands are present in Fig. \ref{fig4}a compared to the simple case in Ref. \onlinecite{trocha}. Also, the maximum value of the figure of merit is substantially reduced in our case, $ZT_{max}\sim 1.6$.

\section{Conclusions}

In conclusion, we have analyzed the thermoelectric properties of the T-shaped double quantum dot  system in the Coulomb blockade regime characterized by nonzero onsite interaction terms in both the main and side dot of the system. Our analysis focused on the so called slow detector limit when the interaction between the conduction electrons from the leads and the localized electrons in the main quantum dot is relatively small compared to the inter-dot interaction, $\Delta/t<1$. In this regime, due to a relatively strong interaction between localized electrons in the main and side dots, the transport properties of the system are strongly influenced by the presence of the side dot. As previously discussed in the literature, the main characteristic of the T-shaped double quantum dot  system is the presence of two transport channels, one involving only the system's main dot and the other involving both the main and the side dots of the system. The interference of the electronic waves transmitted through both channels is responsible for the occurrence of the Fano effect.

We analyzed the system's electrical conduction for different onsite Coulomb values in the two component dots. We found that in the slow detector limit, the electrical  conduction is characterized by two resonances when the energy of the localized level in the system's side dot matches either the Fermi energy of the conduction electrons in the leads or the value of the onsite Coulomb interaction in the side dot. When the onsite Coulomb interaction in the side dot is very large ($U_2\rightarrow\infty$) one of the resonances is pushed at very low energies and in this case only one resonance can be observed experimentally. We also analyzed the system's thermal conductivity and we found a similar behavior, however, the structure of the two resonances in this case is richer. When we combine the electric and thermal conductivities of the system we found a strong deviation from the standard Wiedemann-Franz law, in particular in the region of the two Fano resonances. Finally, we considered the system Seebeck coefficient. As a general characteristic, the Seebeck coefficient can take both positive (electron conduction) and negative (hole conduction) values, and cancels at values of the energy level in the side dot corresponding to the Fano resonances.

Using the  thermoelectrical parameters we calculated the system's figure of merit $ZT$. As a general rule, in standard materials the figure of merit is relatively small, $ZT<1$. Here, we found that in T-shaped double quantum dot  systems $ZT$ can be considerably increased even in the presence of strong onsite Coulomb interaction. In particular, one can tune the figure of merit of the system by varying the energy level in the system's side dot until it matches the values for the Fano resonances. We found that $ZT$ can be also tuned by varying the interaction between the conduction electrons in the leads and the localized electrons in the system's main dot. For a value as low as $\Delta/t=0.1$ we predict a maximum value of $ZT\simeq 4$ when the temperature is of the order of $k_B T/t=0.1$. In the vicinity of the Fano resonances ($\ve_2/t=-0.2$ and $\ve_2/t=-2.4$) the system's figure of merit presents maxima around temperature values of the order of $k_BT/t\sim 0.1$. However, these maximum values are significantly reduced in the presence of onsite Coulomb interaction compared with the simple case without onsite Coulomb interaction.

\begin{acknowledgments}
IT would like to acknowledge financial support from the Incentive Grant program at CSUF. 
\end{acknowledgments}

\end{document}